%% file: BZ_Jets_and_Cosmic_Strings_PRD_Revised.tex
\documentclass[preprint,prd]{revtex4-2}

\usepackage{graphicx}
\usepackage{dcolumn}
\usepackage{bm}
\usepackage{subcaption}
\usepackage{amsmath,amsfonts, amssymb}
\usepackage[colorlinks=true,linkcolor=blue,citecolor=blue,urlcolor=blue]{hyperref}

\begin{document}

\title{Blandford-Znajek Jets and the Total Angular Momentum Evolution of a Black Hole Connected to a Cosmic String}

\author{Ishan Swamy}
\author{Deobrat Singh}
\email{deobrat.singh@mitwpu.edu.in}
\affiliation{Department of Physics, Dr. Vishwanath Karad MIT-World Peace University \\ Kothrud, Pune, India}

\date{\today}

\begin{abstract}
Rotating black holes with strong magnetic fields lead to an outward energy flux in the form of jets governed by the Blandford-Znajek mechanism. These jets depend on factors such as accretion rate, magnetic flux and the spin of the black hole. When such rotating black holes get attached to a cosmic string, it leads to a further rotational energy extraction, leading to a reduced spin. We consider such a system and investigate the effect this reduced spin has on the jet power and its dependence on the cosmic string tension, $\mu$. It is shown that for a constant magnetic flux and accretion rate, the jet energy flux is inversely proportional to $\mu^2$. Interestingly, the rate of this energy flux varies with time and is again dependent on $\mu$. We also study the total angular momentum evolution of the black hole by considering four major effects: accretion, jets, cosmic string energy extraction and the Bardeen-Petterson effect. Further, we attempt to analyse the condition for the spin-down of a black hole due to these effects and find out that it is possible for both small and large string tensions, with a higher possibility for larger string tensions. Another interesting phenomenon that has been proposed is the alignment of the jet with the cosmic string. Additionally, the Bardeen-Petterson effect also leads to alignment or misalignment of the inner and outer disks depending on the alignment of the string. In this manuscript we propose that these results might have an observable effect and hence could serve as a potential detection method for cosmic strings.
\end{abstract}

\maketitle
\section{Introduction}
Black holes are formed when the radius of a collapsing star becomes lesser than its Schwarzchild radius arising from the Schwarzchild solution to the Einstein field equations \cite{Schwarz16}. Black holes although arising from a classical theory of gravity have been studied using semi-classical approaches with \cite{Hawking75}\cite{Page76} discussing the thermodynamics, entropy and particle emission of black holes. Quantum effects of emergent black holes have also been discussed in Schwarzchild de Sitter space \cite{Kapoor14}. \\

When black holes interact with cosmic strings which are one dimensional energy densities formed as a result of topological defects in the early universe \cite{Kibble76}\cite{Vilenkin85}, it leads to fascinating effects \cite{Larsen96}. Cosmic string breaking could lead to pair production of black holes \cite{Hawking95} and at an enhanced rate in presence of a magnetic field \cite {ASHOORIOON21}.\\

Circular strings in black hole spacetime are proposed to oscillate in unstable periodic orbits \cite{Frolov99}\cite{Larsen99}\cite{Dubath07}\cite{larsen1994}. This process is chaotic and leads to the eventual capture or scattering of the string. Further, in the case of charged circular string, stable stationary strings are theorized to exist around the black hole\cite{Larsen94}. Strings, due to their formation in the early universe are proposed to have interactions with primordial black holes, objects that also formed in the early universe \cite{Vilenkin18}. It is suggested that this interaction leads to the formation of large black hole-cosmic string networks.\\

In the case of rotating black holes, strings attached to it leads to possible production of smaller string loops\cite{Xing21}, and additionally can lead to its rotational energy being extracted by the string\cite{Kinoshita16}. This leads to a potential spin-down of the black hole and its mass loss\cite{ahmed24}. Consequently this can lead to indirect observable effects such as the change in the orbital period of the companion star if the black hole is part of an X-ray binary system \cite{Swamy25}.Further if the black hole is accreting, its radius of innermost stable circular orbit depends on the string tension \cite{singh25} and the black hole never reaches maximum spin despite high accretion rates when attached to a cosmic string with a sufficiently large string tension\cite{Xing21}. \\

Studies have been conducted on cosmic strings and their interactions, however there has been no conclusive evidence of its existence yet. Current detection methods include observing their gravitational waves \cite{Vachaspati85} with LIGO-Virgo-KAGRA collaboration data able to put and upper bound of $10^{-11}$ on the cosmic string tension \cite{Abbott21}. Another prominent method is the gravitational lensing effect of the cosmic string \cite{Sazhin03} with a recent study suggesting cosmic strings as pssible candidates for the lensing effect on the galaxy pair SDSSJ110429.61+233150.3 \cite{Safonova24}. \\

We propose here that another observable effect might be the change in the power of the Blandford-Znajek (BZ) jets. These jets are an outward energy flux resulting due to the presence of strong magnetic fields around an accreting black hole\cite{Blandford77}\cite{Juhan02}. These jets extract the black hole’s spin energy and and its rate depends on the dimensionless Kerr spin parameter, $a = J_{BH} /M^2$ of the black hole (in geometric units) where $M$ is the mass of the black hole and $J_{BH}$ is its angular momentum\cite{Kerr63}\cite{Misner73}. \\

In this work, we discuss the energy flux of BZ jets in section \ref{BZ flux} and then analyse its dependence on the cosmic string tension in section \ref{BZ flux cs}. We find out that the energy flux is reduced due to the presence of cosmic string, and the rate of this outward flux depends on the string tension, due to the mass loss by the string. Furthermore, if the string isn't aligned with the black hole spin, it leads to black hole aligning with the string and consequently leads to the jet also aligning with the string which could be potentially observable. Additionally, if the accretion disk is misaligned due to the the Lense-Thirring effect i.e., the Bardeen-Petterson effect \cite{Bardeen75}, the system becomes more complex with alignment depending on the angular momentum vector of the black hole, string and the disk. This can potentially lead to observable effects such as breaking of disk structure into discrete rings due to misalignment \cite{Nealon15}.\\

Further, we consider four major effects: accretion, BZ mechanism, Bardeen-Petterson effect and cosmic string energy extraction, to estimate their effect on the angular momentum evolution of the black hole attached to a string in section \ref{Total J}. In section \ref{Spindown}, we attempt to analyse the conditions for the possibility of spin-down of a black hole in the cases of small and large string tensions. We propose that spin down is possible in both cases, however it is less likely for smaller string tensions.

\section{Effect on Blandford-Znajek Jets} 

\subsection{BZ mechanism}
\label{BZ flux}
Consider a black hole of mass $M$ accreting at a rate $\dot{M}_{acc}$ to have a magnetic field flux $\Phi$ around it, causing the formation of Blandford-Znajek(BZ) Jets \cite{Blandford77}. These jets extract rotational energy of the black hole given by \cite{talbot21},

\begin{equation}
    \dot{E}_{BZ} = \frac{\kappa}{4\pi}\frac{\Phi^2}{M^2}f(a)
    \label{jets}
\end{equation}

where $\kappa$ is a parameter depending weakly on the magnetic field geometry.
 The dimensionless function $f(a)$ of the Kerr spin parameter $a$ is given by,
\begin{equation}
	\begin{aligned}
	f(a) = &\left(\frac{a}{2(1+\sqrt{1 - \alpha^2})}\right)^2 + 1.38\left(\frac{a}            {2(1+\sqrt{1 - \alpha^2})}		\right)^4\\	
       &\times -9.2\left(\frac{a}{2(1+\sqrt{1 - \alpha^2})}\right)^6
	\label{f}
	\end{aligned}
\end{equation}

In the limit $a<<1$, the energy extraction rate becomes, 
\begin{equation}
    \dot{E}_{BZ} = \frac{\kappa}{4\pi}\Phi^2 \frac{a^2}{16M^2}
    \label{jetslim}
\end{equation}

We notice from (\ref{jets})and (\ref{jetslim}) that the energy flux of BZ jets depends only on the magnetic flux, mass of the black hole, and the spin parameter and not on the accretion rate.

\subsection{Cosmic string attached to the black hole}
\label{BZ flux cs}
\begin{figure}[ht!]
	\centering
	\includegraphics[width = \textwidth, height = 7cm]{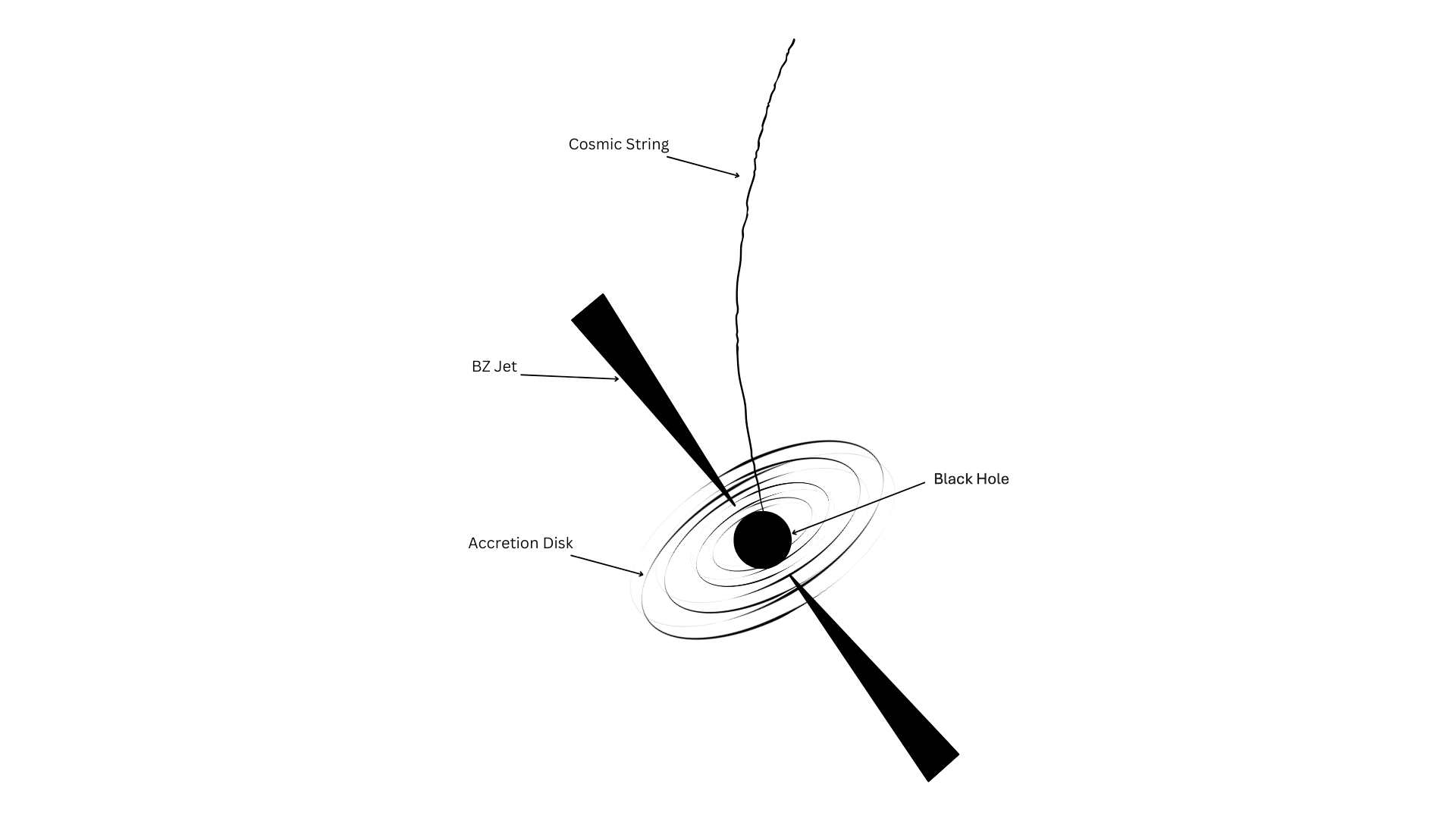}
	\caption{Illustration of the Accreting Black Hole - Cosmic String system}
	\label{diag1}
\end{figure}

We consider a Nambu-Goto cosmic string attached to the accreting black hole. The string has one attached to the black hole and the other end extending outwards as depicted in Fig \ref{diag1}. The invariant length is considered to be extremely large in comparison to the black hole radius. The direct interaction of the jets and the string are not considered in this work. It is to be noted that in this manuscript we consider $G = c = 1$ and the units of masses and mass rates are in $M_\odot$ and $M_\odot/s$ respectively.

The extraction of rotational energy by cosmic string \cite{Kinoshita16} results in a spin down torque on the black hole. For a cosmic string tension $\mu \lesssim \dot{M}_{acc,G}$ (where $\dot{M}_{acc,G}$ is the dimensionless accretion rate in geometric units), the spin-up accretion torque overpowers the spin-down torque\cite{Xing21} which results in the black hole to continue to spin-up till it reaches $a\approx1$. Thus, we can write from (\ref{f}),
\begin{equation}
	 f(a) \approx \left(\frac{1}{2}\right)^2 + 1.38\left(\frac{1}{2}\right)^4 \times -9.2\left(\frac{1}{2}\right)^6
	\approx -0.237
\label{f(1)}
\end{equation}
This gives,
\begin{equation}
    \dot{E}_{BZ} \approx -0.237 \frac{\kappa}{4\pi}\frac{\Phi^2}{M^2}
    \label{jets1}
\end{equation}

Thus, it is shown that for string tensions $\mu \lesssim \dot{M}_{acc,G}$, there is no visible effect of the presence of cosmic string.

However, for the case of $\mu \gtrsim \dot{M}_{acc,G}$, the spin of the black hole saturates at \cite{Xing21}
\begin{equation}
a \sim \dot{M}_{acc,G}/\mu
\label{asat}
\end{equation}
Inserting (\ref{asat}) in (\ref{f}) gives,
\begin{equation}
	\begin{aligned}
	f(a) = &\left(\frac{\dot{M}_{acc,G}/\mu}{2(1+\sqrt{1 - \dot{M}_{acc,G}/\mu^2})}\right)^2 + 1.38\left(\frac{\dot{M}_{acc,G}/\mu}            {2(1+\sqrt{1 - \dot{M}_{acc,G}/\mu^2})}		\right)^4\\	
       &\times -9.2\left(\frac{\dot{M}_{acc,G}/\mu}{2(1+\sqrt{1 - \dot{M}_{acc,G}/\mu^2})}\right)^6
      \end{aligned}
\end{equation}
\begin{equation}
 f(a) = \left(\frac{\dot{M}_{acc,G}}{2(\mu + \sqrt{\mu^2 - \dot{M}_{acc,G}^2})}\right)^2  -12.696\left(\frac{\dot{M}_{acc,G}}{2(\mu + \sqrt{\mu^2 - \dot{M}_{acc,G}^2})}\right)^{10}
\label{fsat}
\end{equation}
Hence $f(a)$ and the corresponding $\dot{E}_{BZ}$ is dependent on the string tension, highlighting the effect of cosmic string attached to the accreting black hole. 

In the case of tensions  $\mu >> \dot{M}_{acc,G}$, the spin $a<<1$. Using (\ref{jetslim}), 
\begin{equation}
    \dot{E}_{BZ} = \frac{\kappa}{4\pi}\Phi^2 \frac{\dot{M}_{acc,G}^2}{16M^2\mu^2}
    \label{jetssat1}
\end{equation}
Thus, for a constant accretion rate and magnetic flux, the energy extraction rate by the BZ process,
\begin{equation}
    \dot{E}_{BZ} \propto \frac{1}{\mu^2}
    \label{jetssat2}
\end{equation}
Once again it is shown that the energy extraction rate is affected by the presence of cosmic string. \\

The extracted rotational energy leads to a mass loss $\dot{M}_{str}$ \cite{Kinoshita16}\cite{ahmed24}. Now, as previously shown in (\ref{jets}), the energy of the jets does not depend on the change in mass of the black hole, indicating that the mass loss due to string does not affect the energy extraction rate. However, the rate of energy flux does depend on both the accretion rate and mass loss by the string as,
 \begin{equation}
 	\begin{gathered}
    	\ddot{E}_{BZ} = -\frac{\kappa}{4\pi}\frac{(\dot{M}_{acc}-\dot{M}_{str})}{M^3}\Phi^2f(a)\\
   	 \Rightarrow  \dot{E}_{BZ}(t) = -\frac{\kappa}{4\pi}\left(\frac{(\dot{M}_{acc}-10^4 \mu)t}{M^3}\right)\Phi^2f(a)
  	  \label{sdjets}
    \end{gathered}	
\end{equation} 
where $\dot{M}_{str} =  10^4 \mu (M_\odot/s)$ is the mass loss by cosmic string \cite{Kinoshita16}\cite{ahmed24}.\\

\begin{figure}[ht!]
\centering
\includegraphics[width=0.75\linewidth, height=8cm]{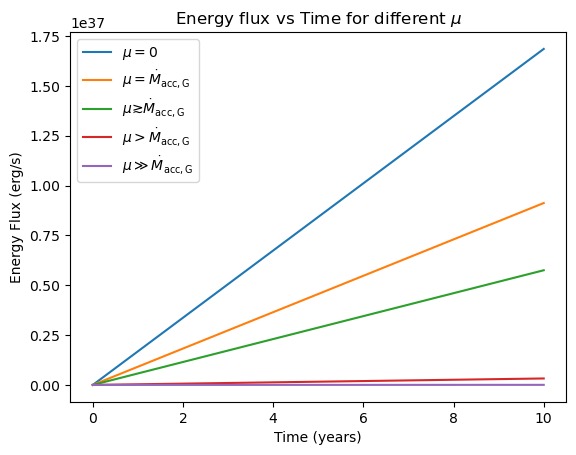} 
\caption{Energy flux vs Time for different string tensions}
\label{fig2}
\end{figure}

We now use (\ref{sdjets}) for the supermassive black hole M87* to examine the variation of energy flux considering scenarios both with and without the influence of cosmic strings characterized by different string tensions. Recent observations by the Event Horizon Telescope \cite{Akiyama19, Akiyama21} place the mass of M87* at $6.5 \times 10^9 M_{\odot}$, with the magnetic field strength estimated between $1$ and $30$ G, and the accretion rate $\dot{M}_{acc}$ ranging from $3$ to $20 \times 10^{-4} M_{\odot}\mathrm{yr}^{-1}$. For the purposes of this analysis, we fix values of $B = 10$ G and $\dot{M}_{acc} = 10^{-3} M_{\odot}\mathrm{yr}^{-1}$. We then compute and plot the time evolution of the energy flux, $\dot{E}_{\mathrm{BZ}}(t)$, over a 10-year period for various string tension values $\mu$, as illustrated in Fig. {\ref{fig2}}

The $\mu = 0$ line (depicted in blue) represents the predicted energy flux in the absence of a cosmic string, exhibiting a linear increase with time. In contrast, when a cosmic string is present (shown by the yellow, green, red, and purple lines), the growth rate of the energy flux is noticeably reduced, highlighting the pronounced influence of the string on the BZ jet. Notably, in the regime where $\mu \gg \dot{M}_{\mathrm{acc},G}$ (purple line), the energy flux remains nearly constant throughout the considered period. This suggests that if a cosmic string were to become attached to the supermassive black hole at the present time (t=0), its effect on the BZ jet can potentially be observed in a period of 10 years. Such a signature could thus serve as a potential observational method for detecting cosmic strings.

\section{Total Angular Momentum Evolution of Black Hole Attached to Cosmic String}
\label{Total J}

\subsection{Angular momentum evolution by accretion in the presence of cosmic strings}
\label{accretion}

The angular momentum evolution due to accretion is given by \cite{talbot21}, 
\begin{equation}
\boldsymbol{\dot{J}_{acc}} = \dot{M}_{acc}L_{ISCO}\boldsymbol{j_{BH}}
\label{J_acc}
\end{equation}
The specific angular momentum at the innermost stable circular orbit (ISCO) is 
\begin{equation}
L_{ISCO} = \pm \frac{M}{\Lambda(a)}\frac{\Lambda(a)^2 \mp 2a\sqrt{\Lambda(a)} + a^2}{\sqrt{\Lambda(a) - 3 \pm 2a/\sqrt{\Lambda(a)}}}
\label{L_ISCO}
\end{equation}
where
\begin{equation}
	\begin{gathered}
	\Lambda(a) =  3 + Z_2(a) \mp \sqrt{(3-Z_1(a))(3 + Z_1(a) + 2Z_2(a))}\\ 
	Z_1(a) = 1 + (1 - a^2)^{1/3}[(1 - a)^{1/3} + (1 - a)^{1/3}]\\
	Z_2(a) = \sqrt{3a^2 + Z_1^2(a)}
	\label{lambda}
	\end{gathered}
\end{equation}
with prograde and retrograde motion differentiated by addition and subtraction operators. correspondingly.\\

We now use (\ref{asat}) in (\ref{L_ISCO}) to understand the dependence of $L_{ISCO}$ and consequently $\boldsymbol{\dot{J}_{acc}}$ on the string tension $\mu$. For this, we consider a stellar black hole with $M = 10 M_\odot$ accreting at 3 different rates of $1 M_{Edd}, 0.1 M_{Edd}, 0.01 M_{Edd}$ and $0.001 M_{Edd}$. The corresponding ranges for $\mu$  are taken such that the condition  $\mu \gtrsim \dot{M}_{acc,G}$ is satisfied. We restore the values of $G$ and $c$ here for the simulated data.\\
\begin{figure}[ht!]
\centering
\includegraphics[width=0.75\linewidth, height=9cm]{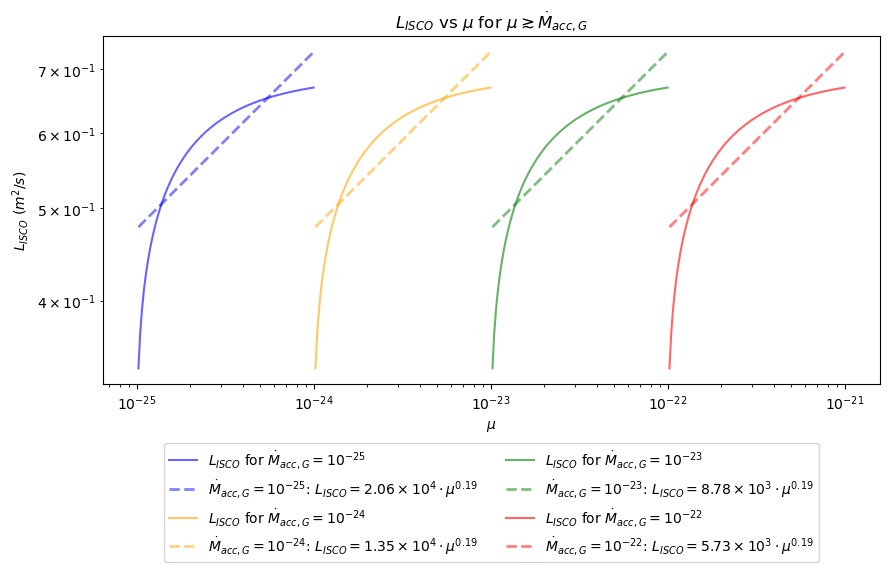} 
\caption{$L_{ISCO}$ vs $\mu$ for $\mu \gtrsim \dot{M}_{acc,G}$}
\label{fig3}
\end{figure}

The plots are fitted with $L_{ISCO} = C \mu^n$ to get the required relation, and from Fig \ref{fig3}, it is estimated that for $\mu \gtrsim \dot{M}_{acc,G}$
\begin{equation}
L_{ISCO} \propto  \mu^{0.19}
\label{L vs mu}
\end{equation}
despite varying accretion rates. Consequently (from (\ref{J_acc})),
\begin{equation}
\dot{J}_{acc} \propto  \mu^{0.19}
\label{J_acc vs mu}
\end{equation}
\begin{figure}[ht!]
\centering
\includegraphics[width=0.75\linewidth, height=9cm]{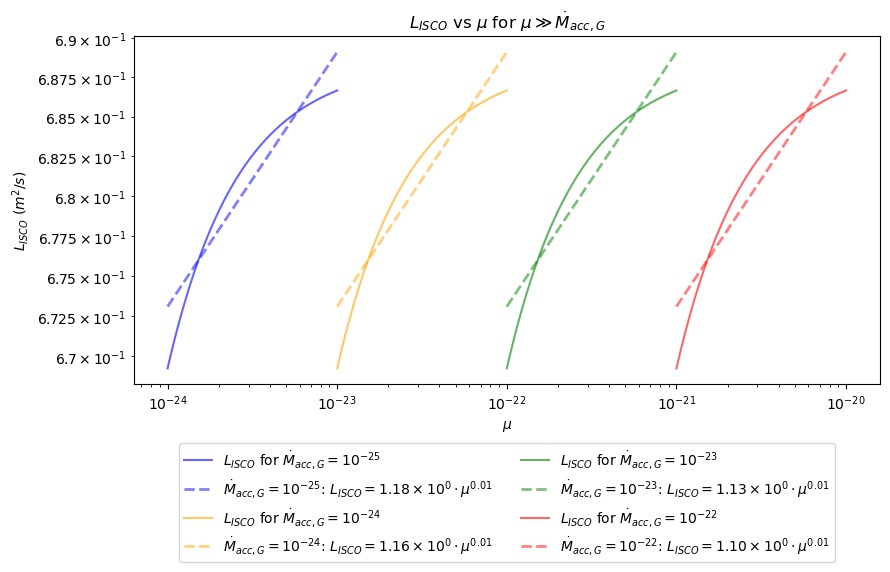} 
\caption{$L_{ISCO}$ vs $\mu$ for $\mu >> \dot{M}_{acc,G}$}
\label{fig4}
\end{figure}
However, it is interesting to note that for $\mu >> \dot{M}_{acc,G}$, as shown Fig \ref{fig4}, 
\begin{align}
L_{ISCO} &\propto  \mu^{0.01} \\
\Rightarrow \dot{J}_{acc} &\propto  \mu^{0.01}
\label{J_acc vs mu large}
\end{align}
i.e., it is nearly independent. Thus, for larger string tensions, the angular momentum evolution due to accretion can be considered constant for varying string tensions.

\subsection{Angular momentum evolution by BZ jets in the presence of cosmic strings}
\label{BZjets}

The angular momentum of the black hole is affected by the BZ jets as \cite{talbot21}, 
\begin{equation}
\boldsymbol{\dot{J}_{BZ}} = -\dot{M}_{acc}\boldsymbol{L_{BZ}}
\label{J_BZ}
\end{equation}
The effective specific angular momentum $\boldsymbol{L_{BZ}}$ is
\begin{equation}
\boldsymbol{L_{BZ}} = \frac{\kappa}{2\pi}\frac{\Phi^2}{M\dot{M}_{acc}} \left(\frac{a}{2(1+\sqrt{1 - \alpha^2})}\right)\boldsymbol{j_{BH}}
\label{L_BZ}
\end{equation}
where $\boldsymbol{j_{BH}}$ is the unit vector along the black hole angular momentum. 

In the presence of cosmic strings, for the case $\mu \lesssim \dot{M}_{acc,G}$, it is easy to see that, like the energy flux, the angular momentum remains unaffected. For $\mu \gtrsim \dot{M}_{acc,G}$, 
\begin{equation}
\boldsymbol{L_{BZ}} = \frac{\kappa}{2\pi}\frac{\Phi^2}{M\dot{M}_{acc}} \left(\frac{\dot{M}_{acc,G}}{2(\mu + \sqrt{\mu^2 - \dot{M}_{acc,G}^2})}\right)\boldsymbol{j_{BH}}
\label{L_BZstr}
\end{equation}
and for  $\mu >> \dot{M}_{acc,G}$,
\begin{equation}
\boldsymbol{L_{BZ}} = \frac{\kappa}{8\pi}\frac{\Phi^2}{M\dot{M}_{acc}}\frac{\dot{M}_{acc,G}}{\mu}
\label{L_BZstr2}
\end{equation}

\subsection{Angular momentum evolution due to cosmic strings and black hole spin alignment}
\label{cosmicstrings}

Now, the black hole angular momentum evolution due to cosmic strings is \cite{Xing21}, 
\begin{equation}
\boldsymbol{\dot{J}_{CS}} = -\frac{\mu}{M}[\boldsymbol{{J}_{BH}} - (\boldsymbol{j_{CS}\cdot J_{BH}})\boldsymbol{j_{CS}}]
\label{J_CS}
\end{equation}
where $\boldsymbol{j_{CS}}$ is the unit vector along the string.

\begin{figure}[ht!]
	\centering
	\includegraphics[width = \textwidth, height = 7cm]{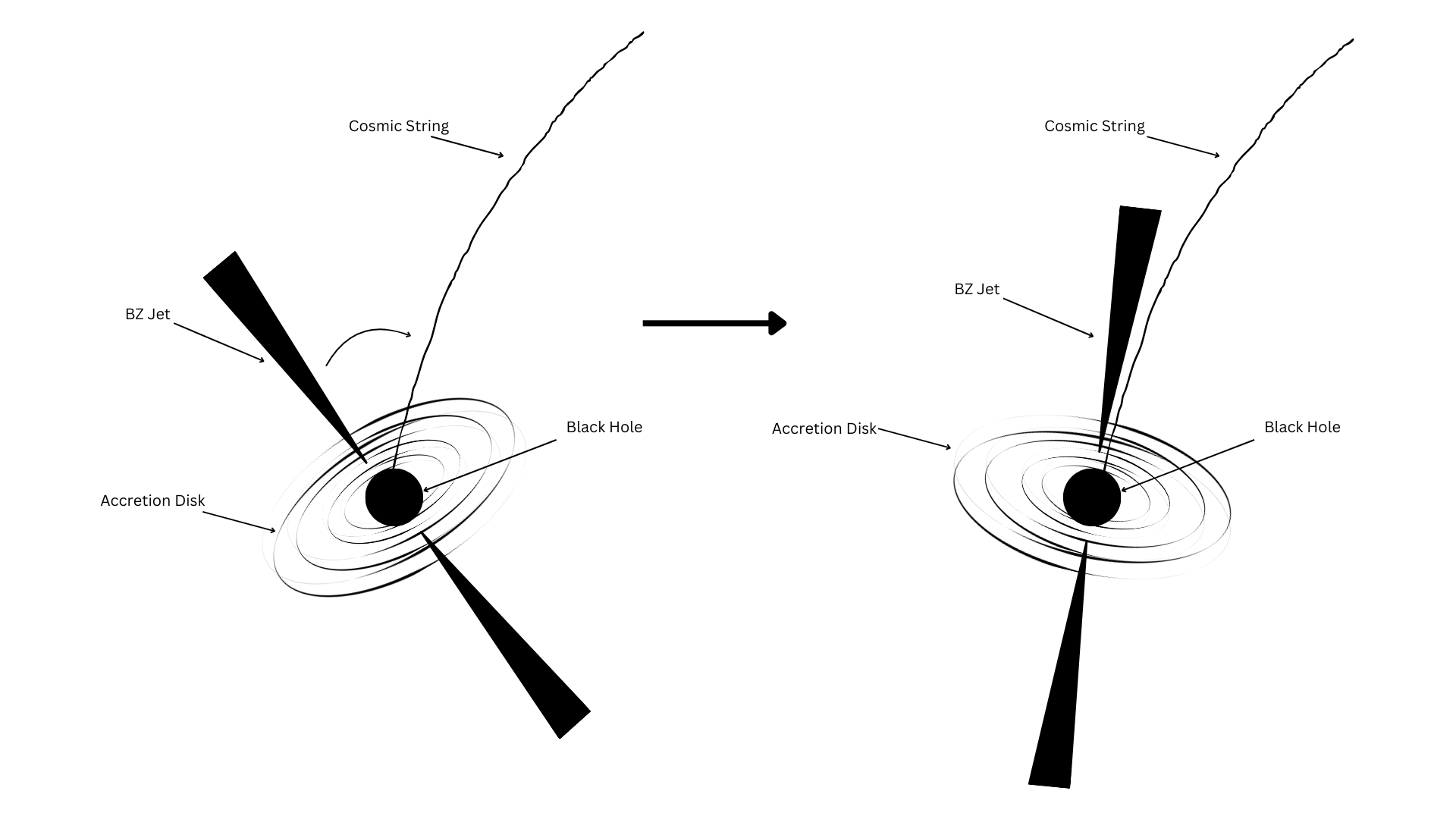}
	\caption{The black hole and its BZ jet aligning with the cosmic string}
	\label{diag2}
\end{figure}

BZ jets are aligned with the angular momentum of the black hole, as shown in (\ref{L_BZ}). However, if the cosmic string is not aligned with the black hole, it results in the black hole's angular momentum aligning itself with the string. From (\ref{J_CS}), the perpendicular component of $\boldsymbol{\dot{J}_{CS}}$ is given by, 
\begin{equation}
\boldsymbol{J^\perp_{CS}} = \boldsymbol{J^{\perp0}_{CS}} exp\left[-\frac{\mu}{M}t\right]
\label{J_CSperp}
\end{equation}
where $t_\mu = M/\mu$ is the timescale for the black hole spin alignment with the cosmic string as in \cite{Xing21}. Hence, it can be inferred that the BZ jet also aligns with the cosmic string (Fig \ref{diag2}). For a low-mass black hole (ex. a stellar black hole) and a cosmic string with large string tension, the timescale becomes smaller, thus providing for potential observational effects. 

\subsection{Bardeen-Petterson effect}
\label{BPeffect}

The Bardeen-Petterson effect for a black hole with angular momentum $J_{BH}$ leads to an angular momentum evolution given by \cite{talbot21}\cite{pringle92}\cite{King05}\cite{Martin07}
\begin{equation}
\boldsymbol{\dot{J}_{BP}} = - \frac{J_{BH}}{\tau_{GM}}\{sin(\pi/7)(\boldsymbol{j_{BH}} \times \boldsymbol{j_d}) + cos(\pi/7)[\boldsymbol{j_{BH}} \times (\boldsymbol{j_{BH}} \times \boldsymbol{j_d})]\}
\label{J_BP}
\end{equation}
with $\bm j_{BH}$ and $\bm j_d$ being unit vectors along black hole angular momentum and disc angular momentum, respectively. The gravitomagnetic time-scale is \cite{Martin07}\cite{Perego09}\cite{Dotti13}
\begin{equation}
\tau_{GM} \approx 0.17 \left(\frac{M}{10^6}\right)^{2/35}\left(\frac{f_{Edd}}{\epsilon_r/0.1}\right)^{-32/35}a^{5/7} Myrs
\label{tau_gm}
\end{equation}
where $f_{EDD}$ is a fraction of Eddington mass used to parameterise $\dot{M}_{acc}$ and $\epsilon_r$ is the spin-dependent radiative efficiency. Now,
\begin{equation}
f_{Edd} \approx 0.76 \left(\frac{\epsilon_r}{0.1}\right)\left(\frac{M_d}{10^4}\right)^{5}\left(\frac{M}{10^6}\right)^{-47/7}\left(\frac{aJ_d}{3J_{BH}}\right)^{-25/7}
\label{f_edd}
\end{equation}
where $M_d$ and $J_d$ are the disc mass and angular momentum, respectively. We insert (\ref{f_edd}) 
and notice that for a given $M, M_d, J_{BH}$ and $J_d$,
\begin{equation}
\tau_{GM} \propto  a^{190/49}
\end{equation}
and consequently,
\begin{equation}
|\dot{J}_{BP}| \propto a^{-190/49}
\label{BP_a}
\end{equation}

Further, we use the previously defined condition $\mu \gtrsim \dot{M}_{acc,G}$ where $a \sim \dot{M}_{acc,G}/\mu$ such that (for a constant accretion rate)
\begin{equation}
|\dot{J}_{BP}| \propto \mu^{190/49} \approx \mu^{3.87}
\label{BP_mu}
\end{equation}
Thus, it is shown that not only does the Bardeen-Petterson angular momentum evolution depend on the string tension, but it is proportional to nearly its fourth power. \\

\begin{figure}[ht!]
	\begin{subfigure}	{0.45\linewidth}
		\includegraphics[width = 1\linewidth, height = 6cm]{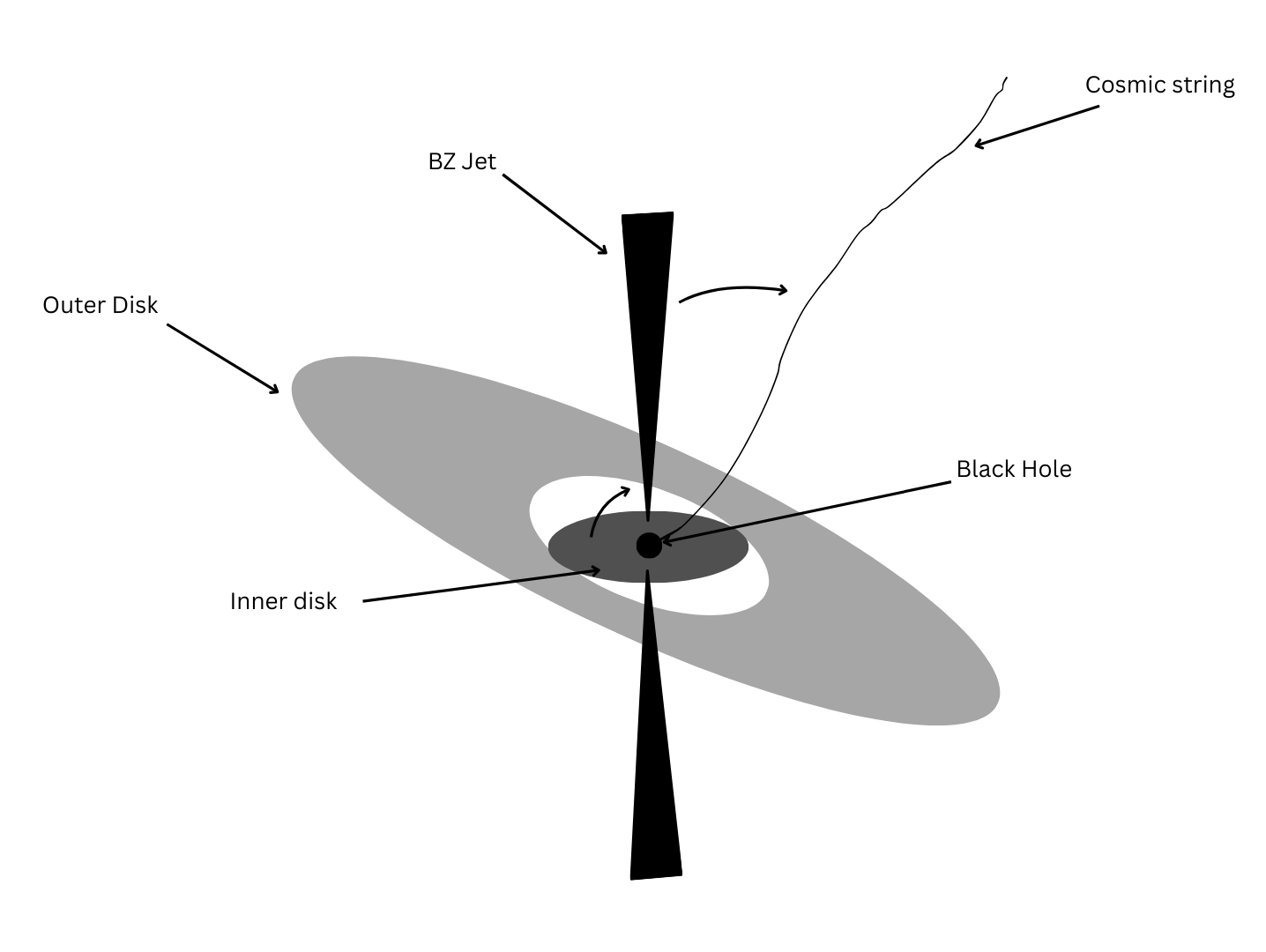}
		\captionsetup{justification = centering}
		\subcaption{Inner and outer disk alignment}
		\label{diag3}
	\end{subfigure}
	\begin{subfigure}	{0.45\linewidth}
		\includegraphics[width = 1\linewidth, height = 6cm]{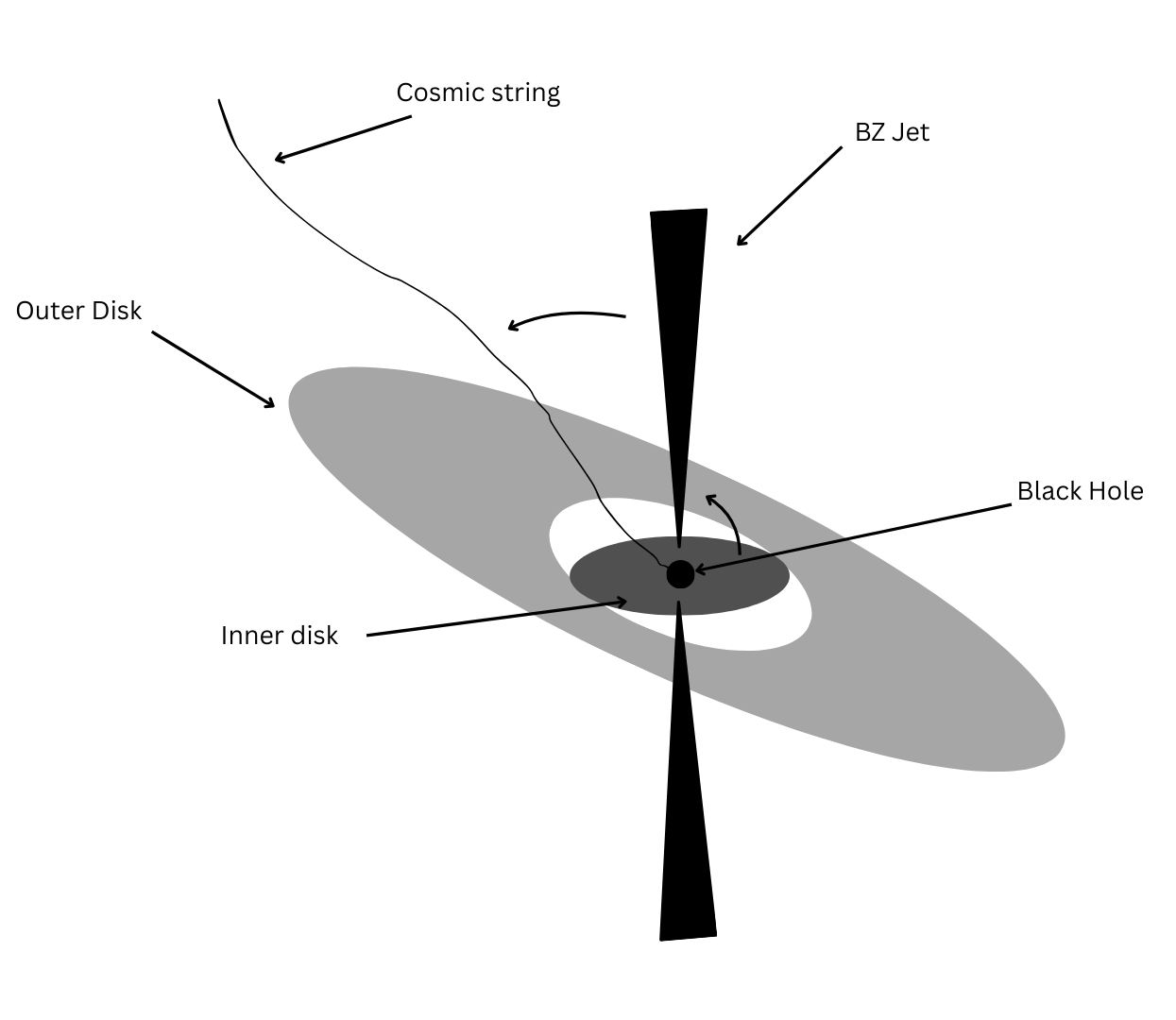}
		\captionsetup{justification = centering}
		\subcaption{Inner and outer disk misalignment}
		\label{diag4}
	\end{subfigure}
\caption{Bardeen-Petterson effect in presence of cosmic string}
\label{fig5}
\end{figure}

\section{Spin-Down of a Black Hole}
\label{Spindown}

Now the total angular momentum of black hole evolves due to accretion ($\dot{J}_{acc}$), BZ process ($\dot{J}_{BZ}$), energy extraction by cosmic string ($\dot{J}_{CS}$) and the Bardeen-Petterson effect($\dot{J}_{BP}$). Thus, 
\begin{equation}
\bm
\dot{J}_{BH} = \dot{J}_{acc} + \dot{J}_{BZ} + \dot{J}_{BP} + \dot{J}_{CS}
\label{JTotal}
\end{equation}
The black hole undergoes a spin-down process if 
\begin{equation}
|{\dot{J}_{CS}}|  + |\dot{J}_{BZ}| + |\dot{J}_{BP}| > \dot{J}_{acc}
\label{spcond}
\end{equation} 
It is extremely important to note that the values of $\mu$ are generally in the order of negative powers of ten \cite{Turok86} such as the upper bound of $10^{-11}$on string tension set by the LIGO - Virgo - KAGRA collaboration \cite{Abbott21}. Using this value in our results of eq. (\ref{J_acc vs mu}) from section \ref{accretion}, it can be roughly established that 
\begin{equation}
	\begin{aligned}
	\dot{J}_{acc} &\propto  \mu^{0.19} \sim 10^{-3} (\mu \gtrsim \dot{M}_{acc,G})\\
	\dot{J}_{acc} &\propto  \mu^{0.01} \sim 10^{-1} (\mu >> \dot{M}_{acc,G})
	\label{acc_cont}
	\end{aligned}
\end{equation}
Further from eq. (\ref{L_BZstr}) and eq. (\ref{L_BZstr2}) from section \ref{BZjets}, eq. (\ref{J_CS}) from section \ref{cosmicstrings}, and eq. (\ref{BP_mu}) from section \ref{BPeffect}, we get
\begin{equation}
	\begin{aligned}
	|\dot{J}_{BZ}| & \propto \mu^{-1} \sim 10^{11}\\
	|\dot{J}_{CS}| &\propto \mu \sim 10^{-11}\\
	|\dot{J}_{BP}| &\propto \mu^{190/49} \sim 10^{-43}
	\label{contri}
	\end{aligned}
\end{equation}
Hence, from eq. (\ref{contri}) it can be infered that $|\dot{J}_{BZ}|$ becomes the dominating term contributing in the spin-down of the black hole in comparison to the other two terms. \\

Combining these results, it can be shown that that the spin-down condition of eq. (\ref{spcond}) might be satisfied for larger string tensions even if the black hole has high accretion rates. For smaller string tensions, the lower contributions of $|{\dot{J}_{CS}}|$ and $|\dot{J}_{BP}|$ might lead to the condition not being satisfied unless the accretion is extremely low.  However, the extremely high contribution from  $|{\dot{J}_{BZ}}|$ could once again lead to the condition being satisfied and hence cause the spin-down of the black hole. It is important to point out here that the exact constraints on the string tension and spin-down condition require a detailed investigation.

\section{Conclusion}
In this manuscript, we focused on the effect of a cosmic string attached to the black hole on the BZ jets, specifically its outward energy flux. It is shown that for smaller string tensions ($\mu \lesssim \dot{M}_{acc,G}$) the outward energy flux remains the same even in the presence of cosmic string (\ref{jets1}), whereas for larger string tensions ($\mu \gtrsim \dot{M}_{acc,G}$), the energy flux is reduced in the presence of a cosmic string in compared to the flux in its absence (\ref{fsat}). Further, for very large string tensions, the energy flux becomes inversely proportional to the square of the string tension for a constant magnetic flux and accretion rate (\ref{jetssat2}). However, as established in (\ref{sdjets}), the outward energy flux is not constant with time and depends on the difference in accretion rate and the mass loss by the cosmic string. This is a significant result as this may lead to observable effects and serve as potential evidence for the presence of an attached cosmic string. \\

Further, the evolution of the total angular momentum of a black hole attached to a cosmic string has been investigated, by considering four significant effects: accretion, BZ jets, cosmic string energy extraction, and the Bardeen-Petterson effect. It is noted that the effect of cosmic string on the angular momentum due to accretion is large for $\mu \gtrsim \dot{M}_{acc,G}$ (eq. (\ref{J_acc vs mu})) and becomes significant for $\mu >> \dot{M}_{acc,G}$. This is an interesting result as it can be inferred that the angular momentum of a black hole heavily depends on the accretion momentum in the presence of cosmic string. We further estimate the effect of the cosmic string on the other three effects and find out that the angular momentum evolution due to BZ jets is inversely proportional to the string tension (eq. (\ref{L_BZstr}) and eq. (\ref{L_BZstr2})), whereas the angular momentum due to energy extraction by the cosmic string is directly proportional to the string tension (eq. (\ref{J_CS}))\cite{Xing21}. The Bardeen-Petterson effect, however, varies as nearly the fourth power of string tension (\ref{BP_mu}), thus establishing that it is the lowest contributor to the total angular momentum of the black hole in comparison to the other three effects(eq. (\ref{contri})). We then combine all these effects to analyse the spin-down condition of a black hole (eq. (\ref{spcond})) and find out that it is highly possible in the case of large string tensions. Even for small string tensions, spin-down might be possible, but the likelihood and the exact constraints require a detailed investigation. \\

Another interesting phenomenon that has been discussed is the alignment of the black hole spin with the cosmic string. Due to the alignment of the BZ jets with the black hole spin, this leads to the BZ jets also aligning with the string potentially leading to observable effects, especially for stellar mass black holes. This could serve as another detection method for the presence of cosmic strings. Furthermore, the presence of the Bardeen-Petterson effect might result in slower or faster disk alignment with the black hole, depending on the alignment of the string (Fig \ref{fig5}). \\

We would like to mention again that the cosmic string considered in this manuscript is uncharged. However, in the case of a charged string, the magnetic field of the black hole and the BZ jets could lead to potential electromagnetic and thermal effects resulting in possible breaking and decaying of strings. Moreover, oscillating strings could also lead to additional effects. These require further analysis and will be carried out by the authors in the future.  

\section{Acknowledgement}
The authors acknowledge Ms. Niyukti patil for her contribution in Fig \ref{diag1}, Fig \ref{diag2} and Fig \ref{fig5}.

\input{BZ_Jets_and_Cosmic_Strings_PRD_Revised.bbl}

\end{document}

%% file: BZ_Jets_and_Cosmic_Strings_PRD_Revised.bbl
%